\documentstyle[11pt,paspconf,epsfig]{article}

\begin{document}

\title{Numerical models of rotating accretion flows around black holes}

\author{Igor V. Igumenshchev\altaffilmark{1}}
\affil{Astronomy \& Astrophysics, G\"oteborg University and Chalmers
    University of Technology, 412 96 G\"oteborg, Sweden}


\altaffiltext{1}{Present address: Institute of Astronomy, 48 Pyatnitskaya
Street, 109017 Moscow, Russia}


\begin{abstract}
Numerical, two-dimensional, time-dependent
hydrodynamical models of geometrically thick
accretion discs around black holes are presented.
Accretion flows with non-effective radiation cooling (ADAFs)
can be both convectively stable or unstable depending on the
value of the viscosity parameter $\alpha$.
The high viscosity flows ($\alpha\simeq 1$) are stable
and have a strong equatorial inflow and bipolar outflows.
The low viscosity flows ($\alpha\la0.1$)
are convectively unstable and this induces quasi-periodic variability.
\end{abstract}




\section{Introduction}

Advection-dominated accretion flows (ADAFs)
have recently attracted much attention
since they naturally explain the properties of
X-ray transients, low luminosity active galactic nuclei and
other high energy objects.
Most of the information of ADAF structure derives from a simplified
vertically integrated approach,
which reduces the complicated three-dimensional problem
of accretion flow hydrodynamics to a one-dimensional problem
(see Narayan \& Yi 1995;
Chen, Abramowicz \& Lasota 1997;
Narayan, Kato \& Honma 1997; Igumenshchev, Abramowicz \& Novikov 1998).
In the vertically integrated approach only the radial structure of the
disc is studied in a detailed way.
Due to
significant simplifications introduced by the vertical integration,
some important effects, such as convection
(Narayan \& Yi 1994; Igumenshchev, Chen \& Abramowicz 1996) or
outflows (see discussion in Narayan \& Yi 1994) are
not properly treated.
Understanding of ADAF properties could be improved by a discussion
of two-dimensional time-dependent hydrodynamical models,
where one explicitly
treats both the radial and vertical structure of the accretion flow.

In this contribution we present preliminary results
of two-dimensional numerical simulations of rotating accretion flows 
around black holes. Complete discussion of the results will be
published (Igumenshchev \& Abramowicz 1999).

We construct hydrodynamical time-dependent models with
viscosity parameter $\alpha\sim 0.1-1$
and high geometrical thickness.
We assume a simplified model of the radiative cooling:
$(1-\varepsilon)$ of the energy generated by  the viscous dissipation
is radiated away. 
We consider a large value of $\varepsilon$
($0.5\leq\varepsilon\leq 1$), when the
accretion flow has a low efficiency of the conversion of its internal
energy to the escaping radiation.
We demonstrate that stability of the flow strongly depends on the value
of viscosity. The low viscosity flows
($\alpha\la 0.1$) are convectively unstable,
and the instability produces a quasi-periodic behaviour of
the accretion flows and outflows.
In the case of high viscosity ($\alpha\sim 1$), the convective
instability is suppressed, and the flow is stable.

\section{Numerical method}

Our numerical technique is based on the solution of
the non-relativistic  Navier-Stokes equations
in the spherical coordinate system ($r$, $\theta$, $\varphi$)
in a stationary gravitational field of the black hole.
The gravity of the black hole is modeled by the
Newtonian potential
\begin{equation}
\Phi(r)=-{c^2\over 2}{r_g\over r},
\end{equation}
where $r_g=2GM/c^2$ is the gravitational radius of black hole of
mass $M$.
We assume axial symmetry
of the flow with respect to the rotational
axis that coincides with the $\theta=0$ direction.
We take into account the contribution of all components of the
viscous stress tensor to the equations of motion and the energy equation.
Shear viscosity is only considered.

For simplicity,
we describe the dynamics of accreting plasma in the framework of
the one fluid approximation.
We use the equation of state for an ideal gas.
The adiabatic index of gas is assumed to be a constant, and takes the value
$\gamma=3/2$. In this case
the equation of thermal energy conservation can be written in the
form: 
\begin{equation}
\rho T{dS\over dt}=Q_{visc}-Q_{rad},
\end{equation}
where the operation $d/dt$ is the comoving (Lagrangian) time derivative,
$S$ is the specific entropy, $Q_{visc}$ is the dissipation function
and $Q_{rad}$ is the volume cooling rate.
The problem of the radiation losses in the high temperature magnetized
plasma is a quite difficult one and it is far from the complete solution.
We do not address this problem here, but instead assume a simple model
for the cooling rate,
\begin{equation}
Q_{rad}=(1-\varepsilon)Q_{visc},
\end{equation}
where $\varepsilon$ is a parameter, $\varepsilon\leq 1$.
The case $\varepsilon=1$ corresponds to the non-radiating
accretion flow, whereas
$\varepsilon=0$ corresponds to the isentropic flow.

The kinematic viscosity coefficient is described by
\begin{equation}
\nu=\alpha{c_s^2\over \Omega_K},
\end{equation}
where $\alpha$ is a constant, $0<\alpha\la 1$,
$c_s= \sqrt{RT/\mu}$ is
the isothermal sound speed, and $\Omega_K=c\sqrt{r_g/2 r^3}$
is the Keplerian angular velocity.

To solve the Navier-Stokes equations we split the numerical procedure of one
time step $\Delta t$ calculation into two parts, hydrodynamical and
viscous. The hydrodynamical part is calculated by using the
explicit finite-difference PPM algorithm
developed by Colella \& Woodward (1984).
The viscous part is solved by the implicit
operator splitting method 
for the contributions to the equations of motion.
The viscous contribution to the thermal energy equation (2) is calculated
using an explicit scheme.
The time step $\Delta t$ is chosen in accordance with
the Courant condition for the hydrodynamical sub-step.

We use the absorbing inner boundary at the radius $3 r_g$, 
which is the location of the last stable orbit of
the Schwarzschild black hole.
The outer boundary is located at the radius
$\simeq 300 r_g$.
At this radius we inject the matter with a fixed rate $\dot{M}_{inj}$,
close to the equatorial plane.
Injected matter has zero $r$- and $\theta$-components
of velocities, and the angular momentum close to the 
Keplerian one.



We look for stationary or quasi-stationary solutions,
which are obtained in the course of the time-dependent calculations
from the initial state.
This requires to follow the evolution
of flow pattern during a few
characteristic accretion times, measured as an average time of motion of the
matter from the outer boundary to the inner one.

%

\section{Results}

We have calculated a number of evolved models with four different 
values of 
$\alpha=1$, $0.3$, $0.1$ and $0.03$, and different values of $\varepsilon$,
which vary in range $0.5-1$. The models show a weak dependence on
$\varepsilon$, and they strongly depend on $\alpha$.

\begin{figure}
\centerline{\epsfig{file=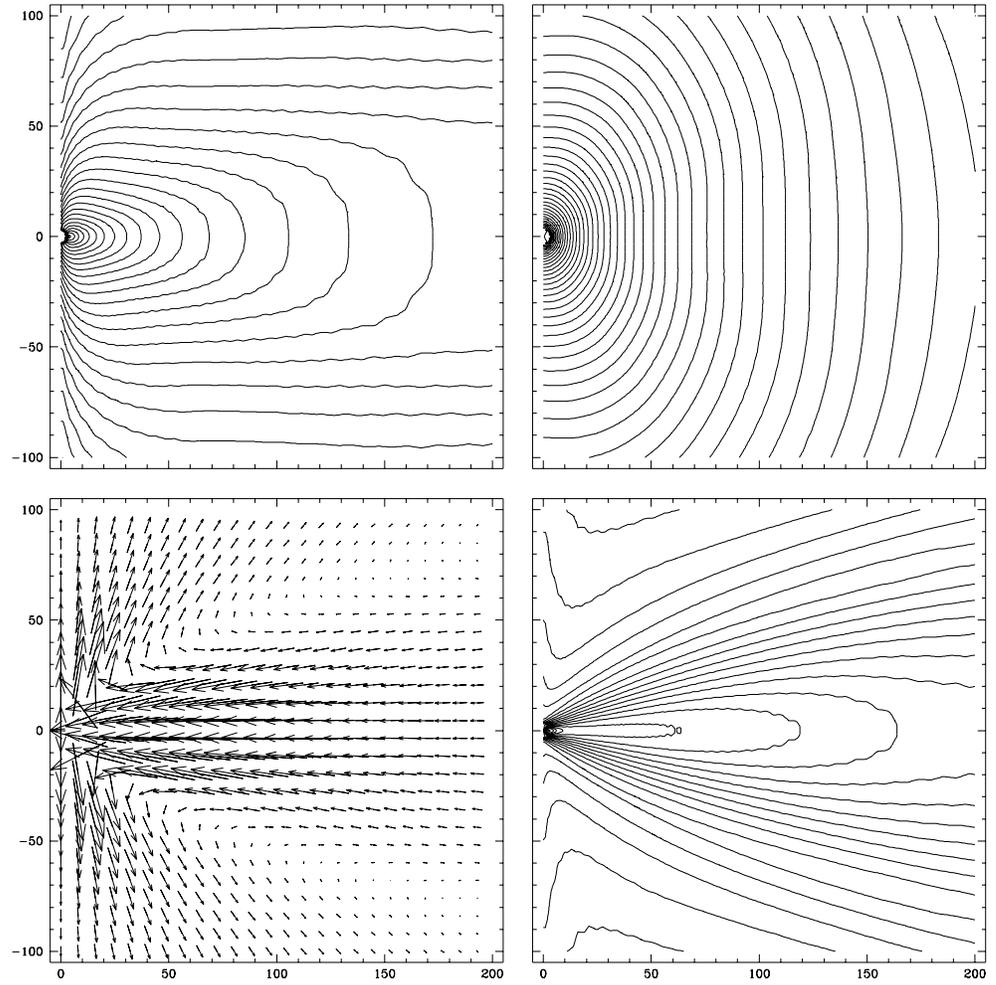,width=13cm,height=13cm}}
\caption {The flow pattern of stable model with $\alpha=1$ and
$\varepsilon=1$ (see text for details).}
\label{fig1}
\end{figure}

For large value of $\alpha=1$ and $0.3$ the models are
stable.
They are symmetric with respect to the equatorial plane
and show no time-dependent behaviour.
Figure~\ref{fig1}
represents the flow pattern of the model with $\alpha=1$ and 
$\varepsilon=1$.
In this figure the meridional cross-section of the model is shown, and
vertical axis coincides with the axis of rotation. Black hole locates
in the origin $(0,0)$. Axes are labeled in the units of $r_g$.
Upper left plot shows the contours of density $\rho$.
The contour lines are spaced with
$\Delta\log\rho=0.1$. 
Upper right plot shows the contours of pressure $P$. The lines are spaced with
$\Delta\log P=0.1$. The density and pressure monotonically increase
toward the black hole.
In lower left plot the arrows with the length
in relative units show the momentum vectors $\rho{\bf v}$.
The flow pattern consists of the equatorial inflow and bipolar outflows,
which originate very close to the black hole, at radius $8 r_g$.
Lower right plot shows the contours of Mach number $\cal M$.
The lines are spaced with $\Delta{\cal M}=0.05$.
The maximum value of $\cal M$ at given radius is reached at the
equatorial plane. Two stagnation points, where ${\cal M}=0$, locate
at the axis of rotation at radius $8 r_g$. The flow is subsonic
(${\cal M}<1$) everywhere
in our computation domain, which has inner boundary at $r=3 r_g$.

The model with $\alpha=0.3$ does not show deep outflows. The stagnation
points locate at radius $80-90 r_g$ on the axis of rotation. Inside
this radius the matter moves to the black hole almost radially.
Oppositely to the case of $\alpha=1$ the distribution of the
Mach number in this model has an equatorial minimum at a given radius.
This minimum indicates that the inflowing matter is overheated
in the inner equatorial parts due to viscous dissipation.
In the low viscosity models this overheating is the reason
for development of the convective instability.

\begin{figure}
\centerline{\epsfig{file=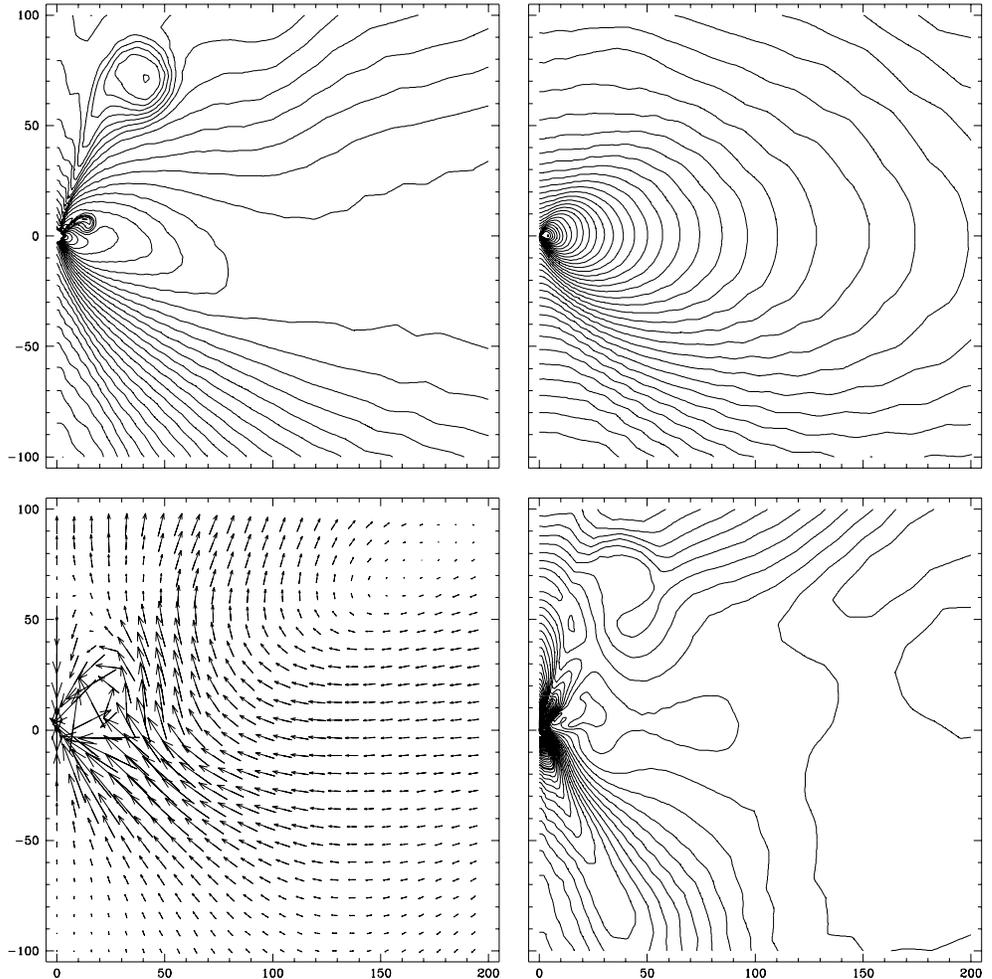,width=13cm,height=13cm}}
\caption {The flow pattern of convectively unstable model with $\alpha=0.1$ and
$\varepsilon=1$ (see text for details).}
\label{fig2}
\end{figure}

Low viscosity models with $\alpha\le 0.1$ are unstable.
They demonstrate rich and complicated time-dependent variations of flow pattern.
Example snapshot of
flow pattern of the convectively unstable 
model with $\alpha=0.1$ and $\varepsilon=1$ is shown in Figure~\ref{fig2}.
In the upper left plot for density distribution
it is clearly seen the reason for the instability --- hot convective
bubbles, which quasi-periodically originate in the innermost region
($4-6 r_g$ from the center)
of accretion flow and propagate outward.

The bubbles arise from the
initial perturbations, which usually appear
in the $\theta$-directions, where the maximum mass flux
onto the black hole occurs at that moment.
As a rule, the bubbles originate near,
slightly above or below, the equatorial plane.
When a bubble has developed, it forces
the direction of the maximum mass inflow to change.
In the new direction, a new convective bubble originates
when the previous bubble reaches the radial distance 50--100 $r_g$.
This cycle repeats quasi-periodically.
Typically, convective instability produces sequences
of convective bubbles, where
the previous and next bubbles originate and move outward in different
hemispheres with respect to the equatorial plane.
The quasi-periodic behaviour of the convective bubbles,
accompanied by the outflows
in polar directions, produces
a significant variability of the flow pattern inside $r\sim 100 r_g$.
The intensity and directions (up or down)
of the outflows strongly correlate with
the convective activity in the vicinity of the black hole.

The spatial scale of perturbed motion in the
accretion flow becomes smaller
with decreasing of the viscosity
parameter $\alpha$. Accordingly, the flow pattern is more
complicated in model with  $\alpha=0.03$.
Small scale vortices accompany the convective motion.
The vortices exist quite a long time and
strongly interacts with  convective bubbles.

We have studied the time-dependent behaviour of the unstable models
calculating
radiative energy losses in the case of
proton-electron bremsstrahlung cooling for optically thin plasma,
\begin{equation}
 L(t) \propto \int\rho^2 T^{1/2}dV.
\end{equation}
Here the integration is taken
over the volume of a sphere with the radius $100\; r_g$.
Fourier analysis of time series of the radiative cooling rates
shows the presence of 
a strong feature at frequencies $10-20\; (M_\odot/M)\; {\rm Hz}$
in the power spectrum of model with $\alpha=0.1$.
In the case of $\alpha=0.03$, a weaker feature
is observed at $1-2\; (M_\odot/M)\;{\rm Hz}$.
These oscillations 
can be explained
by quasi-periodic presence of convective bubbles and perturbations
of the accretion flow introduced by these bubbles.
Note, this variability has typical time scales in the observed
quasi-periodic oscillations
(QPOs) range of the Galactic black hole candidate X-ray sources.

\end{document}